\documentclass[prd,twocolumn,showpacs]{revtex4}

\usepackage{multirow}
\usepackage{rotating}
\usepackage{amssymb}
\newcommand{\dd}{\mathrm{d}}

\newcommand{\omM}{\Omega_{\mathrm{M}}}
\newcommand{\omB}{\Omega_{\mathrm{b}}}
\newcommand{\omK}{\Omega_{K}}
\newcommand{\omL}{\Omega_{\Lambda}}

\begin{document}
 
\title[Rogues' Gallery]{Rogues' gallery: the full freedom of the Bianchi CMB anomalies}

\author{Andrew Pontzen}
\email{apontzen@ast.cam.ac.uk}
\address{Institute of Astronomy, Madingley Road, Cambridge CB3 0HA, UK }
\date{14 January 2009}

\begin{abstract}
  Combining a recent derivation of the CMB evolution equations for
  homogeneous but anisotropic (Bianchi) cosmologies with an account of
  the full linearized dynamical freedoms available in such models, I
  calculate and discuss the various temperature and polarisation
  anisotropy patterns which may be formed. Certain anisotropies can be
  hidden in superhorizon modes at early times, thus avoiding any
  constraints from nucleosynthesis while nevertheless producing
  non-trivial redshift-zero temperature patterns in flat and open
  universes. The results are likely to be more of pedagogical than
  observational interest, but future work will assess whether such
  patterns can be matched to anomalies in WMAP results.
\end{abstract}

\maketitle

\section{Introduction}

Bianchi universes -- a large and almost complete class of relativistic
models which are homogeneous but not necessarily isotropic -- have
remained of interest to mathematical cosmologists since seminal papers
uncovering their major properties in the 1950s and 60s
\cite[e.g.][]{EllMac69,estabrook1968das,heckmann1962gic,taub1951est};
for a concise and accessible review see
Ref. \cite{2006GReGr..38.1003E}.  However for observers they are often
considered obsolete, the chaotic cosmology programme having been
superseded by the inflationary paradigm, especially in the light of
CMB temperature maps from COBE \cite{1992ApJ...396L...1S} and WMAP
\cite{2008arXiv0803.0732H} which are statistically very close to
isotropy.  On the other hand, there are deviations from the standard
model predictions, consistent between COBE and WMAP, which are not
easily linked to known systematics
(e.g. \cite{2005PhRvL..95g1301L,2006PhRvD..74b3005D,2007PhRvD..75b3507C}
and references therein). Jaffe~et~al have shown
\cite{2005ApJ...629L...1J} that anomalies such as the low quadrupole
amplitude, alignment of low-$l$ modes and large-scale power asymmetry
can be mimicked by adding (to the standard, scale-free perturbations
which dominate the CMB signal) Bianchi temperature maps derived for a
specific subcase of VII$_h$ by Hawking and later Barrow
\cite{1973MNRAS.162..307C,1985MNRAS.213..917B} (see also
Ref. \cite{1997PhLA..233..169B} in which the temperature patterns are
discussed qualitatively by considering the Killing motions in a local
orthonormal frame). Such models are more predictive than simple
Bianchi I models which can trivially be fine-tuned to solve the
quadrupole problem,
e.g. \cite{2006PhRvL..97m1302C,2008PhRvD..77b3534R,2008ApJ...679....1K}.
Although the cosmological parameters implied by Jaffe et al's VII$_h$
fit are inconsistent with concordance parameters
\cite{2006ApJ...644..701J,2007MNRAS.377.1473B} and constraints from
primordial nucleosynthesis \cite{1976MNRAS.175..359B}, there is a
possibility that these tensions may be resolved by a more complete
dynamical model
\cite{1997PhRvD..55.7451B,2006ApJ...644..701J,2007MNRAS.380.1387P}.

Recently, in Ref. \cite{2007MNRAS.380.1387P} (henceforth PC07), a
complete and computationally convenient Boltzmann hierarchy framework
for calculating temperature and polarisation anisotropies in any
nearly Friedmann-Robertson-Walker (FRW) Bianchi universe was
described. For the favoured parameters based on Jaffe et al's analysis
of observed temperature maps, the predicted $B$-mode polarisation is
too strong to be consistent with observational upper limits from WMAP
three year data \cite{2007ApJS..170..335P}. However, although the
theoretical analysis was general, in numerical results PC07 followed
previous Bianchi CMB studies in employing dynamical solutions for
which the shear decays as $\sigma \propto a^{-3}$, where $a$ is the
FRW scale factor. It is as a direct result of this rapid decay that
the polarisation amplitude is generally very strong relative to the
anomalous temperature anisotropies; thus, like nucleosynthesis and
density parameter inconsistencies, the polarisation conflict may be
resolved by considering more general dynamics.

In this note, I discuss the appearance of the Bianchi CMB in the
generalised nearly-isotropic regime. (Models which are highly
anisotropic at any point after recombination are ruled out by
observations if one assumes the Copernican principle
\cite{1995ApJ...443....1S}.)  I will show that the aforementioned
polarisation and nucleosynthesis conflicts are simultaneously, yet
naturally, avoided in a certain class of models. By fitting these
models to CMB anomalies, constraints on cosmological parameters can be
produced and tested against trusted concordance values; however, this
requires a detailed statistical framework and is postponed to later
work.

The setup and dynamics of the general case are described in \S
\ref{sec:dynamics} and applied to flat and open models in \S
\ref{sec:results:-open-flat}. For pedagogical interest, I give results
from models with a variety of cosmological parameters which are not
necessarily consistent with other constraints. I describe the results
for a closed universe in \S \ref{sec:closed-models} although the
anomalous CMB contribution in such cases is always a quadrupole and
thus less interesting from a phenomenological standpoint. Finally, I
summarise the work in \S \ref{sec:conclusions}.

\section{Set-up and Dynamics}\label{sec:dynamics}

During an epoch of near-isotropy, the dynamics of any Bianchi model
can be analyzed by a decomposition into an isotropic FRW background
and particular linear perturbations which break the isotropy while
maintaining homogeneity with respect to three simply-transitive
Killing vector fields (KVFs).  For example, one might take an
expanding flat model and perturb the expansion rates along one or two
of the perpendicular coordinate axes; this canonical example arises as
a special case (Type~I) in the formalism described below.

The geometrical setup is described by the structure constants $C^k_{ij}$ which
fix the commutators of the three preferred Killing fields $\vec{\xi}_i$: 
\begin{equation}
\left[ \vec{\xi}_i, \vec{\xi}_j \right] = C^k_{ij} \vec{\xi}_k \textrm{ .}
\end{equation}
The explicit antisymmetry $C^k_{ij} = - C^k_{ji}$ and the Jacobi
identities $C^a_{[bc} C^d_{e]a}=0$ together substantially restrict the
possible values for the $C^k_{ij}$; in fact, up to arbitrary scalings
and rotations only four of the initial $27$ values are required to
describe a given space (canonically \cite{EllMac69} these are known as
$n_1, n_2, n_3$ and $a_1$, where $a_1$ should be sharply distinguished
from the FRW scale factor $a$).  The categorization of Bianchi models
into distinct types rests on classifying the inequivalent sets of
these four constants; here I will consider only the small subset which
admit an FRW limit. Succinct but explicit derivations of the relations
between structure constants and the resulting isotropic background
cosmologies are available in PC07; or for more detail see
e.g. Refs. \cite{wainwright1997dsc,ellis1998cmc,PC08}.

I will first consider the open and flat models (Bianchi type VII$_h$
and its limiting types I, V and VII$_0$); these possess two distinct
geometric freedoms.  The first corresponds to a characteristic length
over which the Killing fields spiral relative to a parallel-propagated
frame. In the canonical decomposition the physical scale of this
spiral is fixed by setting $n_1=0$, $n_2=n_3=1$ and scaling the
present horizon size via the dimensionless Hubble parameter, usually
denoted $x$ \cite{1973MNRAS.162..307C,1969MNRAS.142..129H}. The second
freedom corresponds to the isotropic curvature scale and can be set by
specifying the fiducial FRW parameter $\omK$, in terms of which
$a_1=\pm x \sqrt{\omK}$. This sign ambiguity reflects a parity
freedom; we will adopt the conventions of PC07 noting the standard
result that, in terms of the CMB, parity inversion can be achieved by
mirroring all maps and inverting the sign of $B$ polarisation.

As $x\to\infty$ for fixed $\omK$ one pushes spiral structure out of
the horizon and obtains Type V models (this situation can be
renormalized so that $x=1/\sqrt{\omK}$, $n_2=n_3=0$ and $a_1=\pm 1$);
for $\omK=0$ we have the flat Type VII$_0$ specialisation
($n_2=n_3=1$, $a_1=0$); with $x\to\infty$ and $\omK=0$ one obtains
Type I (renormalisable such that $x$ takes an arbitrary finite value
with all structure constants vanishing), which was described at the
start of this section.

Except in type I, the choice of canonical values preselects a
preferred axis, rotations about which leave the structure constants
invariant; the alignment of the perturbed shear eigenvectors relative
to this axis determine both the patterns obtained in the CMB and the
nature of the dynamics, but rotations around the axis are pure gauge.
Thus anisotropic perturbations are divided into three classes which
transform respectively as scalars (henceforth $s$-modes), vectors
($v$-modes) and rank-2 tensors ($t$-modes) about the locally preferred
axis; modes in these classes must evolve independently in the linear
approximation \footnote{This decomposition can be made in the
  non-linear case \cite{2005CQGra..22..579C} and is essentially an
  example of the automorphism approach to Bianchi dynamics
  (e.g. \cite{1988PhR...166...89R} and references therein).}. The
modes generate in turn $m=0,1$ and $2$ spherical harmonics in the CMB
if the fiducial $\theta=0$ direction is taken along the residual
symmetry axis (as in PC07).  The general nearly-FRW Bianchi CMB
appears as a linear combination of the three resulting patterns,
although here we will study each mode separately.
\begin{figure*}
\begin{center}
\includegraphics[width=0.95\textwidth]{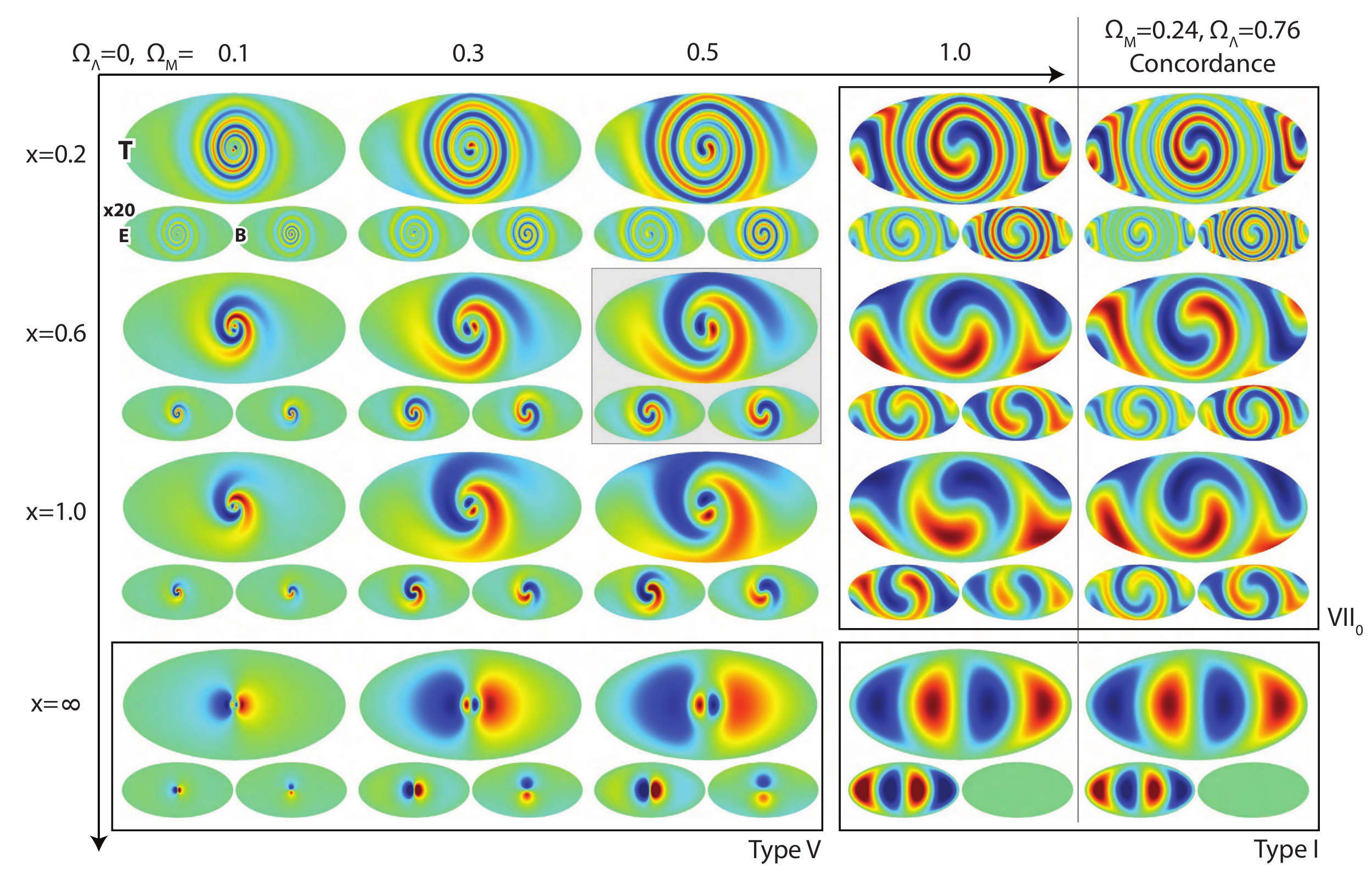}
\end{center}

\caption{The $v$ (vector) modes of type VII$_h$ and its
  specialisations.  For a grid of parameters the temperature pattern
  is shown (upper panel) and $E$ and $B$ mode polarisation exaggerated
  in scale by a factor $20$ (small lower left and right panels
  respectively). $x$ varies from top to bottom while $\omM$ (and hence
  $\omK$) varies from left to right. I assume $\omL=0$ except in the
  rightmost panels for which the concordance values
  ($\omM,\omL=0.24,0.76$) are used. The primary effect of non-zero
  $\Lambda$ is to increase the conformal time to last scattering,
  causing the final $z=0$ pattern to be more tightly wound. When
  $\omM+\omL=1$ one obtains type VII$_0$; as $x\to \infty$ one obtains
  type V (from VII$_h$) or I (from VII$_0$).  The model with
  $(\omM,x)=(0.5,0.6)$ (shaded panel) is able to mimic known CMB
  temperature anomalies
  \cite{2005ApJ...629L...1J,2007MNRAS.377.1473B}.}\label{fig:v-modes}
\end{figure*}

A careful analysis of previous works
\cite{1973ApJ...180..317C,1975JETP...42..943G} shows that the linear
amplitude $A$ of any Bianchi mode evolves according to
\begin{equation}
A'' + 2 \mathcal{H} A' + \lambda A= 0 \label{eq:evol-eq}
\end{equation}
where primes denote the derivative with respect to conformal time
$\eta$, $\mathcal{H}=a'/a$ is the conformal Hubble parameter,
$\lambda$ is a time-independent complex eigenvalue associated with
each decoupled mode and the shear resulting from the mode has
magnitude $\sigma \propto \left| A'\right| /a$.  (When $\lambda \notin
\mathbb{R}$ the argument of the complex-valued $A$ defines the
orientation in the preferred plane; as previously stated the evolution
equations are invariant under rotations in this plane, $A\to
Ae^{im\Delta \phi}$.)  Eq.  (\ref{eq:evol-eq}) assumes there to be no
anisotropic matter source terms, which in the absence of exotic
effects should be a good approximation for the recombination era
onwards (even if neutrino viscosity has an important role to play
during radiation domination
\cite[e.g.][]{1995PhRvD..51.3113B,1997PhRvD..55.7451B}).

Further technical discussion of the Bianchi linearisation and its
relation to the existing literature will be given in a future work
\cite{PC08}; however, its essential features can be understood without
the extensive framework required for a full derivation.  For instance
one may show that each mode maps onto a specific first order FRW
metric perturbation (e.g. Bardeen \cite{1980PhRvD..22.1882B}) with $A$
proportional to the mode amplitude in the natural synchronous gauge
defined by the Bianchi-homogeneous space slices.  Seen from this
perspective, the decaying modes correspond to vector perturbations
(eq. 4.12 -- 4.13 of Ref. \cite{1980PhRvD..22.1882B}, with Bardeen's
$\Psi\equiv A'$) and infinite wavelength gravitational waves (eq. 4.14
of Ref. \cite{1980PhRvD..22.1882B}, with $k\to 0$ and Bardeen's
$H\equiv A$) whereas the oscillating modes correspond to gravitational
waves of specific finite wavelengths (with our $\lambda\equiv k^2+2K$,
where $K$ is the FRW curvature parameter and $k$ is the wavenumber in
Ref. \cite{1980PhRvD..22.1882B}).  The particular wavelengths are
associated, in the open and flat cases, with the aforementioned spiral
lengthscale.  Alternatively one may discern the physical status of the
modes by considering gauge invariant covariant variables
\cite{ellis1998cmc}, in terms of which the oscilliatory modes arise
from the shear propagation equation (\cite{ellis1998cmc}, eq. 31) as
an interaction between shear and anisotropic 3-curvature ($S \propto
|\lambda A|/a^2$).

For closed (Type IX) models the canonical structure constants read
$n_1=n_2=n_3=1$, $a_1=0$; in these units, the Hubble parameter is
fixed by the FRW curvature such that $x=1/\sqrt{-4\, \omK}$, i.e.
$K=1/4$. If the physical curvature radius is made progressively
larger, we must take $\omK\to 0$ and $x\to \infty$. As before this
limit yields, as expected, Type I models (explicitly by renormalizing
such that $n_1=n_2=n_3=0$ and $x$ takes an arbitrary finite value). Further
discussion of the closed models is deferred to \S
\ref{sec:closed-models}.

In all cases the physical value of $H_0$ enters only through the
recombination history which is obtained using \textsc{Recfast}
\cite{1999ApJ...523L...1S} assuming $H_0 = 72 \, \mathrm{km\,s}^{-1}\,
\mathrm{Mpc}^{-1}$ and $\omB=0.044$, but allowing $\omM$ and $\omL$ to
vary consistently with the Bianchi model. For quantitative results
in this pedagogic exploration I have ignored reionisation, discussing
the effects of such an assumption individually for each mode.

\begin{figure*}
\begin{center}
\includegraphics[width=0.95\textwidth]{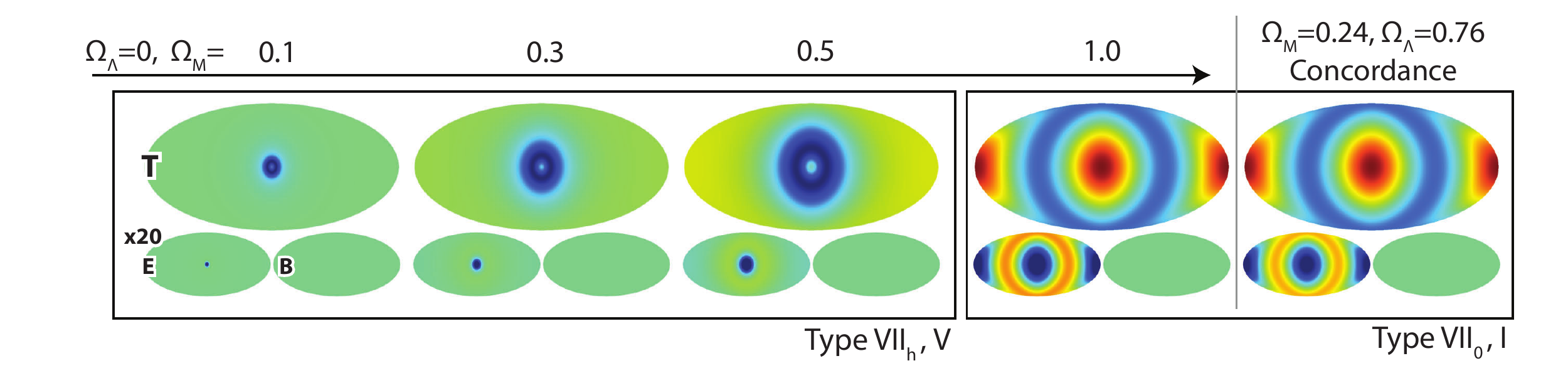}
\end{center}
\caption{The $s$ (scalar) modes of all open and flat types are
  insensitive to the spiralling of the basis vectors, controlled by
  $x$, because the quadrupole anisotropy is aligned along the symmetry
  axis ($m=0$ in the decomposition of PC07). Similarly, no $B$-modes
  can be generated. In the VII$_0$/I limit, there is very little
  sensitivity to the value of $\omL$ since the anisotropy is always a
  pure quadrupole; only the relative polarisation strength is
  marginally affected (because of the increased conformal time to last
  scattering for larger $\omL$). }\label{fig:a-modes}
\end{figure*}

\section{Results: Open and Flat Models}\label{sec:results:-open-flat}

In this section, I will discuss the possible patterns arising in the
CMB when breaking the isotropy of open and flat FRW models; the
relevant Bianchi type is VII$_h$ (together with its previously
described limiting cases). Of the three classes of linearized
anisotropic freedoms discussed above it is the $v$ modes [with
  $\lambda=0$ in Eq. (\ref{eq:evol-eq})] which have been investigated
in previous works on the Bianchi CMB
\cite{2005ApJ...629L...1J,2006ApJ...644..701J,2007MNRAS.377.1473B,2007MNRAS.380.1387P};
a grid of such models for varying $x$ (vertical) and $\omM$
(horizontal) is shown in Fig.~\ref{fig:v-modes}. A similar plot can
be found in Ref. \cite{2006ApJ...643..616J} but here I include the
limiting types (V, VII$_0$ and I) and additionally display the $E$ and
$B$-mode polarisation in two small panels beneath each temperature
map. The model with $(\omM,x)=(0.5,0.6)$ (shaded in
Fig. \ref{fig:v-modes}) generates maps which mimic known CMB anomalies
when suitably oriented on the sky
\cite{2005ApJ...629L...1J,2006MNRAS.367.1714L,2007MNRAS.377.1473B}. The
colour scale of the temperature plot is arbitrary (since one can
rescale the linear perturbation to obtain any specified small
magnitude) but the polarisation panels are shown exaggerated relative
to the temperature panels by a factor of $20$. Because the
polarisation is already strong, reionisation would increase its
strength only marginally (see PC07 for details).

While most models are calculated assuming $\Lambda=0$, the final
column displays concordance $(\omM, \omL=0.24,0.76)$ maps. The
effect of non-zero $\Lambda$ can be understood by examining
equation (33) of PC07: the final patterns obtained are determined by
$a_1$ and the conformal time $\eta$ to last scattering. Keeping these
two quantities constant recovers the $(\omM,\omL,x)$ degeneracy noted
in previous works \cite{2006ApJ...644..701J,2007MNRAS.377.1473B}
(although in the general case this picture is complicated somewhat by
differing $\sigma(a)$ behaviour when $\lambda \ne 0$).

The VII$_h$$s$ mode (Fig. \ref{fig:a-modes}) is similar to the $v$
modes in its dynamical behaviour (also having curvature eigenvalue
$\lambda=0$) and hence produces similar polarisation amplitudes.
Because the mode is invariant under rotations about the preferred
axis, the familiar spiral structure is missing (and hence the value of
$x$ has no effect). Further, $B$-mode polarisation is not generated
(this is necessarily so when the transformation of the mode under
reflections leaving the symmetry axis invariant is considered).  The
only remaining effect is the deformation of the quadrupole into a
focussed spot, which cannot be evaded in anisotropic, homogeneous open
universes \cite{1997PhLA..233..169B,PC08}.

Of more interest are the VII$_h$$t$ modes, which are dynamically
non-trivial with curvature eigenvalue $\lambda=4(1-i a_1)$ in Eq.
(\ref{eq:evol-eq}). The resulting evolution is obtained numerically;
however analytic approximations are helpful in interpreting the
results. For instance during matter domination, while the isotropic
curvature remains on superhorizon scales, we may assume $\omK=0$ and
$\mathcal{H}=2/\eta$, yielding the explicit solution
\begin{equation}
A(\eta) \hspace{-1mm} = \hspace{-1mm}  A_0  \hspace{-1mm} \left[\frac{\cos2\kappa\eta}{2 \left(\kappa\eta\right)^3}  +  \frac{\sin 2\kappa\eta}{\left(\kappa\eta\right)^2} \right] \hspace{-1mm} 
+ \hspace{-1mm}  A_1 \hspace{-1mm}\left[\frac{\sin 2\kappa\eta}{2 \left(\kappa\eta\right)^3} - \frac{\cos 2\kappa\eta}{\left(\kappa\eta\right)^2} \right] \label{eq:matter-sol}
\end{equation}
where the constant $\kappa=\sqrt{\lambda}/2=\sqrt{1-ia_1}$ but
physical results are unaffected by assuming $\kappa \simeq
1$ \footnote{During matter domination, $\eta\left|a_1\right| \ll 1$;
  if $a_1\ll 1$, the approximation $\kappa\simeq 1$ follows
  immediately whereas if $a_1 \sim \mathcal{O}(1)$ one may nonetheless
  approximate $\kappa\simeq 1$ in all physical results by a suitable
  redefinition of the integration constant $A_0 \to A_0/(1-i
  a_1)^{3/2}$ (determined by a Laurent expansion of $\sigma$ and $S$
  in $\eta \ll 1$); the situation is, as expected, physically
  indistinguishable from the flat VII$_0$ case ($a_1=0$).}. $A_0$ and
$A_1$ are integration constants determined by boundary conditions,
e.g. the shear and anisotropic curvature at a fixed time.

Modes with $A_0=0$, $A_1\ne 0$ (`$t_1$ modes') are regular as $\eta
\to 0$ (i.e. their dimensionless shear $\sigma/H=|A'|/\mathcal{H}$ and
anisotropic curvature $S/H^2=|\lambda A|/\mathcal{H}^2$ both tend to
zero); conversely modes with $A_0 \ne 0$ (`$t_0$ modes') have
divergent anisotropy for $\eta\to 0$, thus predicting an early
non-linear phase (which is not itself captured by our linear
description). Although the picture is somewhat complicated by neutrino
free-streaming, matching across matter-radiation equality shows that
one may construct a $t_1$ mode with small shear all the way to the
initial singularity \footnote{Claims that the anisotropic curvature
  causes the shear to diverge logarithmically towards the initial
  singularity, e.g.  \cite{1973JETP...37..739D,1995PhRvD..51.3113B},
  rely on time-averaging over multiple oscillation periods; this is
  meaningful only in models where $\eta \gg 1$ at matter-radiation
  equality, demanding $x\ll \sqrt{\Omega_{\mathrm{R}}}/\omM \sim 0.03$
  for which case the linearisation (\ref{eq:evol-eq}) is inapplicable;
  see \cite{PC08} for more details. Even with the logarithmic
  divergence the nucleosynthesis constraints are drastically weakened
  over power-law divergences
  \cite{1976MNRAS.175..359B,1997PhRvD..55.7451B}.}. Thus by choosing a
late-time solution with $A_0=0$ one will automatically satisfy
nucleosynthesis constraints (see also
\cite{2007PhRvD..75l3517D}). Furthermore, if the amplitude of all
modes is set by some form of equipartition theorem at high redshifts
(e.g. \cite{1995PhRvD..51.3113B}), one may expect on entering the
linear regime that the divergent $t_0$, $v$ and $s$ modes have similar
amplitudes and decay rapidly to leave only these regular $t_1$ modes.
(However numerical models show that, with arbitrary initial
conditions, the amplitude of the $t_1$ mode is often sufficient to
trigger a late-time non-linear anisotropic phase when it enters the
horizon, e.g. \cite{1998CQGra..15..331W}, giving rise to the
intermediate isotropisation picture in which homogeneous shear is only
small for a finite period \cite{1973JETP...37..739D}. Of course if
$\Lambda\ne 0$, this final non-linear phase can be avoided by
shrinking the horizon again at late times.) One should bear in mind
that $t_0$ modes will generate similar CMB patterns but have radically
different dynamical behaviour; however, I will not consider such a
case quantitatively.

\begin{figure*}
\begin{center}
\includegraphics[width=0.95\textwidth]{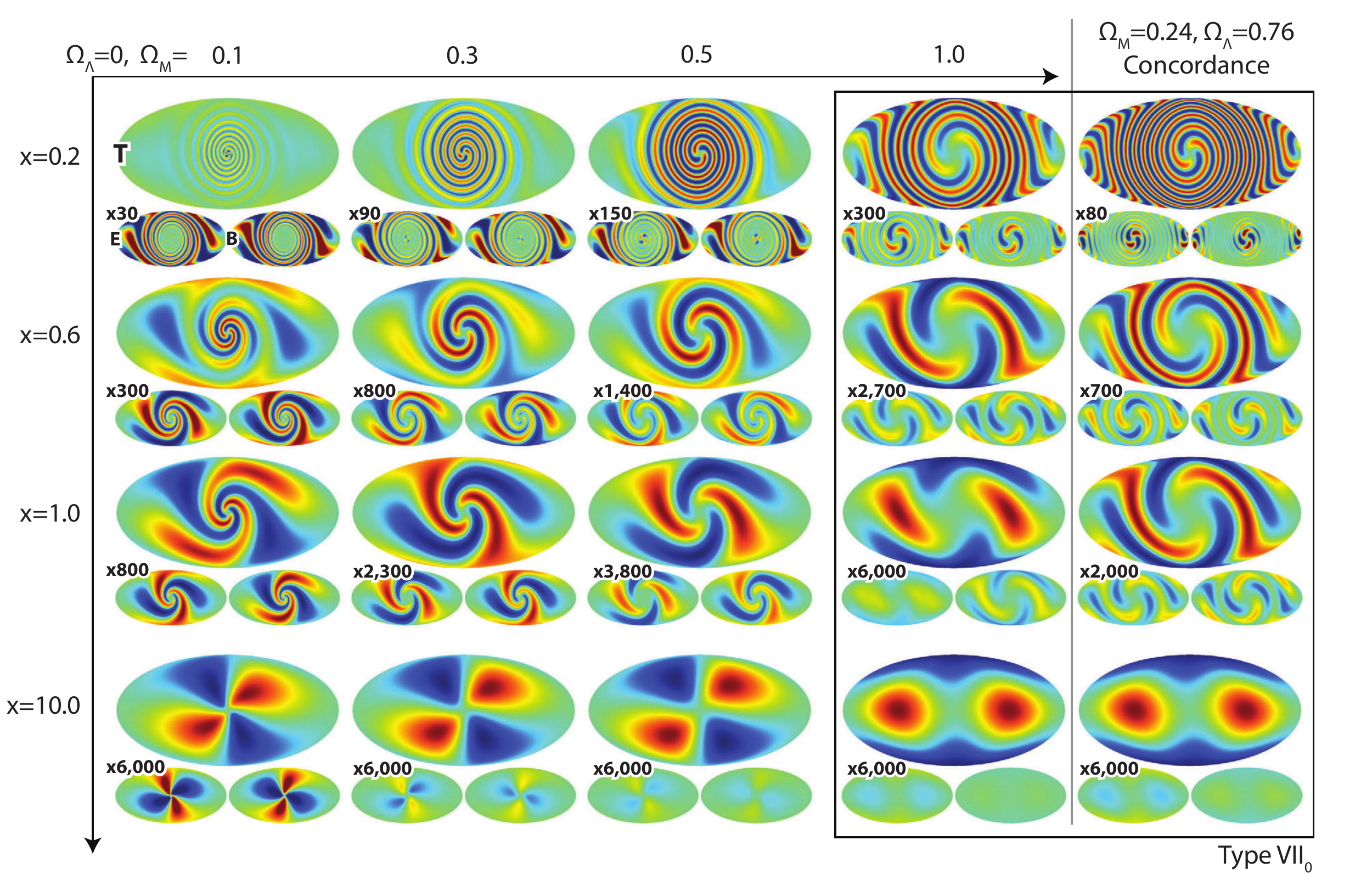}
\end{center}
\vspace{-0.5cm}
\caption{The nucleosynthesis-compatible $t_1$ modes of types VII$_h$
  and VII$_0$. The polarisation magnitude is exaggerated in scale by
  the factors specified in each subplot. The primary polarisation
  varies according to Eq. (\ref{eq:QUT}) and, because the shear can
  grow after recombination, is generally much weaker relative to the
  temperature anisotropy than with $s$ and $v$ modes for which the
  shear always decays.}
\label{fig:t-modes}
\end{figure*} 

The patterns formed by the $t_1$ mode are shown for a grid of
parameters in Fig. \ref{fig:t-modes}. As $x \to \infty$ the
dimensionless curvature eigenvalue $\lambda/x^2$ shrinks to zero (and
thus the anisotropic curvature becomes pure gauge) -- so no strict
Type V or Type I limits exist for $A_0 = 0$ and only finite values of
$x$ are considered. As before, the polarisation shown is purely
primordial and ignores reionisation. The large variations in its
strength between models can be understood by considering the shear
near the decoupling surface. For models where $\omM$ is not tiny,
effects are controlled by the long matter-dominated phase during
which, unaware of small $\Lambda$ or superhorizon isotropic curvature,
observers see a flat Universe with effective Hubble parameter
$\sqrt{\omM} x$. For the polarisation strength one approximates
\begin{equation}
Q,U \propto \sigma_{\mathrm{LSS}}\Delta t \propto \frac{A_1  }{f_ex^2 \sqrt{\omM} H_0 \omB}\label{eq:LSS-shear}
\end{equation}
where $\Delta t$ is the expected time between penultimate and final
scattering, $H_0$ is the physical Hubble parameter and $f_e$ is the
electron fraction at last scattering which scales approximately as
$\omM^{1/2} \omB^{-1} H_0^{-1}$
\cite[e.g.][]{1999coph.book.....P}; since $H_0$ and $\omB$ have been
kept constant for all models, the overall polarisation anisotropy is
taken to be proportional to $A_1 x^{-2} \omM^{-1}$.  On the other
hand, the total temperature anisotropy is built up between
$z_{\mathrm{LSS}}<z<0$. Ignoring the advection of the observed pattern
to estimate the magnitude of the temperature anisotropy we have
\begin{equation}
\Delta T \propto  \int_{t_{\mathrm{LSS}}}^{t_0} \sigma \, \dd t =  \int_{\eta_{\mathrm{LSS}}}^{\eta_0} A' \dd \eta = A(\eta_0) - A(\eta_{\mathrm{LSS}}) 
\end{equation}
where $\eta_0=2/\left(x\sqrt{\omM}\right)$ is the present day equivalent conformal
  time. Inserting the solution (\ref{eq:matter-sol}) one obtains the
  limits
\begin{equation}
\Delta T = A_1 \left\{\begin{array}{ll}
8 \eta_0^2/15 &  \eta_0 \ll 1 \\
4/3 & \eta_0 \gg 1 \textrm{.}
\end{array}\right. \label{eq:deltaT}
\end{equation}
 The final polarisation-to-temperature ratio, combining Eqs
 (\ref{eq:LSS-shear}) and (\ref{eq:deltaT}), should scale roughly as
\begin{equation}
\frac{Q,U}{\Delta T} \propto \max\left(1,\frac{8}{5 x^2 \omM}\right)\label{eq:QUT}
\end{equation}
which agrees with the trends shown in Fig. \ref{fig:t-modes}.  Unlike
the $v$, $s$ and $t_0$ modes, here inclusion of reionisation will
significantly boost the polarisation strength over the primordial
expectation since the majority of models possess a considerably
stronger temperature than polarisation quadrupole at $z<10$.

An obvious question concerns the existence or otherwise of
nucleosynthesis-compatible models which, in a similar vein to the
original VII$_hv$ fits (Fig. \ref{fig:v-modes}, shaded panel), can
mimic known residual CMB anomalies. If this were possible using
parameters consistent with concordance values, i.e. with maps from the
right-hand column of Fig.  \ref{fig:t-modes}, the type VII Bianchi
models would command renewed observational interest.  This will be
investigated, comparing the maps developed here with data from WMAP
\cite{2008arXiv0803.0732H}, in future work.

\section{Comment: Closed Models}\label{sec:closed-models}

Type IX models have been shown by Grishchuk, Doroshkevich and Yudin
\cite{1975JETP...42..943G} to decompose into a closed FRW background
with superimposed maximal wavelength ($k^2 = 6K$) gravitational wave
perturbations (see also Ref.  \cite{1991PhRvD..44.2356K} for helpful
comments) obeying the standard propagation equations
(\cite{1980PhRvD..22.1882B}, eq. 4.14; see end of
\S\ref{sec:dynamics}).  Comparison shows that, within our terminology,
this is modelled by the standard mode evolution (\ref{eq:evol-eq})
with $\lambda=2$ (this result is obtained directly from the
anisotropic field equations in \cite{PC08}).

By an extension of the arguments already presented, setting
$\kappa=\sqrt{2}$ in Eq. (\ref{eq:matter-sol}), one can see that the
primary polarisation is again small for modes which evade
nucleosynthesis constraints; however the temperature patterns are less
phenomenologically interesting as they remain quadrupolar (at first
order). 

To see why this is, consider the origin of non-trivial patterns in the
previously considered open and flat case.  The temperature anisotropy
$\Delta T/T$ arises from integrating the shear tensor $\sigma_{ij}$
projected, on each timeslice, along the tangent vector to the photon
path (PC07, eq.~22).  Taking the
normalized spatial components $p^i$ of the photon momentum (such that
$p^i p_i=1$, $i=1 \cdots 3$), one has
\begin{equation}
\frac{\Delta T(\theta_0, \phi_0)}{T} = -  \int_{t_{\mathrm{LSS}}}^{t_0} \dd t \,
p^i(t; \theta_0, \phi_0) \, p^j(t; \theta_0, \phi_0) \, \sigma_{ij}(t)\label{eq:temp-aniso}
\end{equation}
where the integral is taken between last scattering
($t_{\mathrm{LSS}}$) and the time of observation ($t_0$) and in
general $p^i$ is a function both of observation angles $\theta_0,
\phi_0$ and time $t$.

I have implicitly adopted a spatial chart such that all fluid
quantities have constant components in a given spatial slice (this is
always possible for Bianchi spaces; see PC07).  In particular this
means that the components of the shear tensor $\sigma_{ij}$ depend
only on time $t$, not on the observation angles.  Written in this
basis, non-quadrupolar temperature patterns can only arise through
changes in the components of $\vec{p}$ [at zeroth order for a first
  order solution to Eq. (\ref{eq:temp-aniso})]. In the Type IX case
the components of $p^i$ remain constant to this accuracy \footnote{See
  PC07 eq. 23; in fact it is possible to show that the $p_i$ remain
  constant because the geodesics describe the integral curves of the
  Killing fields, a situation which is also true of the flat Type I
  models but which can never be arranged for open models. This is true
  simply by enumeration of the Bianchi types, but can be seen to arise
  from the divergence of geodesics in hyperbolic spaces
  \cite{WolfPaper}.}; thus the temperature anisotropy remains purely
quadrupolar.

According to the definitions in \S\ref{sec:dynamics}, the shear
oscillates in time such that $\sigma_{ij}(t) = a^{-1}A'(t)
\tilde{\sigma}_{ij}$ for a fixed $\tilde{\sigma}_{ij}$. The
temperature anisotropy (\ref{eq:temp-aniso}) may be written $\Delta
T/T = \left(A(t_0)-A(t_{\mathrm{LSS}})\right)\tilde{\sigma}_{ij} p^i
p^j$ for the fixed observation vector $\vec{p}(\theta_0,\phi_0)$. Thus
the quadrupole intensity oscillates, the microwave background
appearing exactly isotropic when $A(t_0)=A(t_{\mathrm{LSS}})$ which
may occur an indefinite number of times depending on the initial
conditions. Since $x = 1/\sqrt{-4\, \omK}$ (\S \ref{sec:dynamics}),
the physical timescales over which the shear evolves are tied to the
isotropic curvature lengthscale, consistent with the previously noted
interpretation of Type IX anisotropies as maximal wavelength
gravitational waves.

Although the temperature anisotropies remain quadrupolar, $E$ and $B$
modes mix; in fact PC07 eqs (33) or equivalently (50, 52) imply
the polarisation type oscillates with period $\Delta \eta = \pi$. This
effect can be traced to the need for the Killing fields to be
everywhere non-vanishing and yet defined independently of the path
followed between two points, leading to a spiralling of basis vectors
perpendicular to a geodesic (see for example Fig. 1 in
Ref. \cite{1991PhRvD..44.2356K}).

Overall, type IX models can fit almost any combination of quadrupole
anomalies in $T$, $E$ and $B$ modes but, since no $B$ modes can be generated
while the universe remains optically thick, the ratio of the $E$ to $B$
anomalous quadrupole amplitudes determines (in the absence of
reionisation) the conformal time to last scattering and hence
constrains the FRW density parameters through the Hubble
history. Reionisation complicates this picture, as does the existence
of inhomogeneous modes superimposed on the FRW background, but in
principle anomalous quadrupole moments are a powerful constraint for
observers in closed anisotropic universes.

\vspace{-0.3cm}
\section{Conclusions}\label{sec:conclusions}
\vspace{-0.2cm}
I have discussed anomalous CMB signals arising from broken FRW
isotropy in open, flat and closed universes as described by mildly
anisotropic Bianchi models.  These are of interest both pedagogically
and in the context of known CMB anomalies. It is possible to build
anisotropic models which evade traditional constraints from
nucleosynthesis by hiding the anisotropy in super-horizon
gravitational waves which are pure gauge at early times. Whether such
modes can be quantitatively linked to the observed CMB anomalies
remains an open question which will be addressed in a future work.

It is important to stress that all effects discussed in this work,
including for instance $E$-$B$ polarisation mixing, are purely
gravitational and require no magnetic fields, extensions to general
relativity or exotic matter source terms.  Furthermore the models are
predictive -- polarisation-to-temperature ratios and (in the case of
flat and open models) the detailed structure of the possible anomalous
patterns are dictated by FRW density parameters which are constrained
by a range of other observations. Thus a future Bianchi signal
detection yielding consistent cosmological parameters, while highly
surprising from the standpoint of present inflationary frameworks,
would have to be taken seriously.

\section*{Acknowledgments}

I thank Anthony Challinor for collaboration on related work and many
comments on the draft manuscript, Antony Lewis for helpful
discussions, and the anonymous referee for suggestions which improved
the presentation.  Financial support is acknowledged from an STFC
studentship and scholarship at St John's College, Cambridge.

\renewcommand{\apj}{Astrophys. J.~}
\newcommand{\apjl}{Astrophys. J. Lett.~}
\newcommand{\mnras}{Mon. Not. R. Astron. Soc.~}
\newcommand{\apjs}{Astrophys. J. Suppl. Ser.~}
\renewcommand{\prd}{Phys. Rev. D~}
\renewcommand{\prl}{Phys. Rev. Lett.~}
\newcommand{\physrep}{Phys. Rep.~}

\end{document}